\documentclass[prl,twocolumn,eqsecnum,showpacs,preprintnumbers,amssymb]{revtex4}
\usepackage{graphicx}
\usepackage{dcolumn}
\usepackage{bm}
\usepackage{latexsym,epsfig}

\begin{document}

\preprint{\today}

\title{An {\it ab initio} relativistic coupled-cluster theory of dipole and quadrupole polarizabilities: Applications to a few alkali atoms and alkaline earth ions} 
\vspace{0.5cm}

\author{B. K. Sahoo \protect \footnote[2]{E-mail: bijaya@mpipks-dresden.mpg.de}}
\affiliation{Max-Planck Institute for the Physics of Complex Systems \\ N\"othnitzer Stra{\ss}e 38, D-01187 Dresden, Germany}

\date{\today}
\vskip1.0cm

\begin{abstract}
\noindent
We present a general approach within the relativistic coupled-cluster theory framework 
to calculate exactly the first order wave functions due to any rank 
perturbation operators. Using this method, we calculate the static dipole and
quadrupole polarizabilities in some alkali atoms and alkaline earth-metal ions. 
This may be a good test of the present theory for different rank and
parity interaction operators. This shows a wide range of applications including
precise calculations of both parity and CP violating amplitudes due to rank 
zero and rank one weak interaction Hamiltonians. We also give contributions from
correlation effects and discuss them in terms of lower order many-body 
perturbation theory.
\end{abstract} 

\maketitle

For a long time, studies of dipole and quadrupole polarizabilities have been highly interesting in many important aspects for both neutral atoms and ions 
\cite{sternheimer,curtis,vaidyanathan,theodosiou,castro,yan,dzuba,zeng,patil,snow,bromley}. These quantities
are used in the case of ions to remove the quadratic Stark shifts in the recent
proposed optical frequency standards where suitable instruments are not available for very precise measurements \cite{itano,amini}.
Relativistic effects are also investigated for these properties \cite{shuman,lim1}.
Often theoretical studies are limited by many approximations
to only approximate accuracy of the results \cite{kobayashi,porsev}. In many cases, semiempirical methods, which combine the calculated E1 and E2 matrix elements of important states and the experimental excitation energies, are used to calculate
these quantities \cite{safronova,derevianko}.
Core electron contributions are always only estimated in this
approach \cite{sternheimer,safronova}. Therefore, an {\it ab initio} method is necessary to be able to, in principle, explain the
importance of the electron correlation effects in the above properties and also
test the many-body theories employed. These results are largely used to 
determine the van der Waals coefficients or dispersion factors 
\cite{derevianko,porsev,zhu,chu}. In some cases, the Dirac-Fock (DF) method and 
relativistic random-phase approximations (RPA) for the Dalgarno-Lewis \cite{porsev,zhu,dalgarno} or scalar relativistic Douglas-Kroll \cite{lim1,lim2} effective
Hamiltonians are used to determine these quantities. A
number of calculations on these quantities using molecular codes and pseudopotentials have also 
been reported \cite{patil,lim1,lim2,kobayashi}.

Using the relativistic coupled-cluster (RCC) theory, we report here a novel approach for the first time to calculate wave functions up to all orders in the residual Coulomb interaction and first order due to any kind of perturbed operator in 
one valence systems. This takes care of the sum-over-states approach of 
the many-body perturbation theory (MBPT) as an exact solution. This method
can be applied to calculate the first order wave function due to dipole and 
quadrupole transition operators. Hence, polarizabilities can be obtained
 by calculating the expectation values of the corresponding operators.
As an {\it ab initio} test of the theory, it is also possible to calculate the light shift ratio in ions like Ba$^+$ using the present method
as proposed in \cite{sherman}. Moreover, parity non-conserving (PNC)
and CP violating amplitudes can be calculated very precisely using this
method due to rank
zero or rank one weakly interacting Hamiltonian, which are the most challenging 
studies during the last three decades \cite{wood,shukla}. 

 To demonstrate the application of the method, we consider three systems each 
from alkali atoms (Li, Na and K) and alkaline earth ions (Be$^+$, Mg$^+$ and
Ca$^+$) to calculate their dipole and quadrupole polarizabilities having 
different angular momentum selection rules.

The energy shift, $\Delta E(J_n,M_n)$ of any state $|J_nM_n>$, with principal 
quantum number $n$, in a direct current (dc) electric field $\vec \textbf{E}= \mathcal{E} \hat{\textbf{z}}$ can be expressed as
\begin{eqnarray}
\Delta E(J_n,M_n) = -\frac{1}{2} \alpha^1(J_n,M_n) \mathcal{E}^2, 
\end{eqnarray}
where $\alpha^1(J_n,M_n)$ is defined as the static polarizability of state $|J_nM_n>$.
Further, $\alpha^1(J_n,M_n)$ can be divided as
\begin{eqnarray}
\alpha^1(J_n,M_n) = \alpha_0^1(J_n,M_n) + \frac {3M_n^2-J_n(J_n+1)}{J_n(2J_n-1)} \alpha_2^1(J_n,M_n),
\end{eqnarray}
Here $\alpha_0^1(J_n,M_n)$ and $\alpha_2^1(J_n,M_n)$ are known as the scalar and tensor 
polarizabilities, respectively. From the first order perturbation equations,
these parameters can be expressed as the sum over intermediate states
\begin{eqnarray}
\alpha_i^1(J_n,M_n)= -2 \sum_{k \ne n} C_i^1 \frac {|\langle J_nM_n|z|J_kM_k \rangle|^2} {E_n - E_k} ,
\end{eqnarray}
where $i$ represents either $0$ or $2$, $C_i^1$ are the appropriate angular co-efficients, $z$ is the $\hat{\textbf{z}}$ component 
of the position vector $\vec \textbf{r}$ and $E$'s are the unperturbed energy 
levels. Since $z$ can be expressed in terms of the spherical harmonics of rank one 
($Y_{10}(\theta , \phi ))$, the above matrix elements will be non-zero between 
opposite parity states satisfying the E1 transition selection 
rules. The $\alpha_i$'s can be expressed in terms of the reduced matrix elements of
the E1 operator ($D= e \vec \textbf{r}$) as follows
\begin{eqnarray}
\alpha_0^1(J_n) = \frac{-2}{3(2J_n+1)} \sum_{k \ne n} \frac {|\langle J_n||D||J_k \rangle|^2}{E_n - E_k}
\end{eqnarray}
and
\begin{eqnarray}
\alpha_2^1(J_n) = \left (\frac{40J(2J-1)}{3(2J+3)(2J+1)(J+1)}\right )^{1/2} \sum_{k \ne n} (-1)^{J_n+J_k+1} \nonumber \\ \left \{ \matrix { J_n & 1 & J_k \cr 1 & J_n & 2 \cr } \right \} \frac {|\langle J_n||D||J_k \rangle|^2}{E_n - E_k}. \ \ \  \ 
\end{eqnarray}

 Similarly, the static quadrupole polarizability can be expressed as
\begin{eqnarray}
\alpha_0^2(J_n,M_n) &= & -2 \sum_{k \ne n} C_0^2 \frac {|\langle J_nM_n|Q|J_kM_k \rangle|^2} {E_n - E_k} \nonumber \\
 &=& \frac{-2}{5(2J_n+1)} \sum_{k \ne n} \frac {|\langle J_n|| Q ||J_k \rangle|^2}{E_n - E_k} ,
\end{eqnarray}
where $C_0^2$ is the corresponding angular factor and $Q = - \frac{e}{2} (3z^2 - r^2)$ is the E2 operator which has 
different selection rules than the E1 operator.

The above expressions for both the polarizabilities can be expressed in a general form as
\begin{eqnarray}
\alpha(J_nM_n) &=& \langle \Psi_n | O | \Psi_n \rangle \nonumber \\
&=& \langle \Psi_n^{(0)} | O | \Psi_n^{(1)}\rangle + \langle \Psi_n^{(1)} | D | \Psi_n^{(0)}\rangle \nonumber \\
&=& 2 \langle \Psi_n^{(1)} | O | \Psi_n^{(0)}\rangle , 
\label{eqn6}
\end{eqnarray}
where the exact wave function of the $n$th state can be written in terms of the original
atomic wave functions and first order corrections due to the corresponding 
dipole or quadrupole operators $O(=D \ \text{or} \ Q$); i.e.
\begin{eqnarray}
|\Psi_n\rangle &=& |\Psi_n^{(0)}\rangle + |\Psi_n^{(1)}\rangle .
\label{eqn7}
\end{eqnarray}
Note that the $C$ angular factors from the $\alpha$'s are absorbed in the first
order wave functions.

We show that it is possible to calculate $\alpha(J_nM_n)$'s exactly by calculating both 
the $|\Psi_n^{(0)}\rangle$ and $|\Psi_n^{(1)}\rangle$ using single many-body
theory which can avoid the sum-over-states approach given above.
In our approach, we would like to obtain the first order perturbed wave 
function as a solution to the following equation
\begin{eqnarray}
(\text{H}_0^{(\text{DC})} - E_n^{(0)})|\Psi_n^{(1)}\rangle = (E_n^{(1)} - \text{H}_{\text{int}}^{\text{E1}}) |\Psi_n^{(0)}\rangle ,
\label{eqn8}
\end{eqnarray}
where $\text{H}_0^{(\text{DC})}$ and $\text{H}_{\text{int}}$ are 
the Dirac-Coulomb (DC) and interaction Hamiltonians due to E1 or E2
operators, 
respectively. The $E_n^{(0)}$ and $E_n^{(1)}$ are the zeroth and first order
energies of the $n$th state, respectively.

In the RCC theory, we express the exact wave function for one valence ($n$)
state of a system as 
\begin{eqnarray}
|\Psi_n\rangle &=& e^T \{ 1 + S_n \} |\Phi_n\rangle ,
\label{eqn9}
\end{eqnarray}
where $|\Phi_n\rangle$ is the open-shell reference state constructed by appending 
the valence electron '$n$' orbital to the closed-shell Dirac-Hartree-Fock (DF) 
wave function ($|\Phi_0\rangle$; i.e. $|\Phi_n\rangle = a_n^{\dagger} |\Phi_0\rangle$). In the above expression, the coupled-cluster (CC) excitation operators,
$T$ excites only the core electrons and $S_n$  excites either only the valence
electron '$n$' or along with necessary core electrons.

 To get both the unperturbed and perturbed wave functions of Eqn. (\ref{eqn7}),
we express the $T$ and $S_n$ operators by
\begin{eqnarray}
T &=& T^{(0)} + T^{(1)} \label{eqn10} \\
S_n &=& S_n^{(0)} + S_n^{(1)}
\label{eqn11}
\end{eqnarray}
where $T^{(0)}$ and $S_n^{(0)}$ are the CC operators for the DC
Hamiltonian and $T^{(1)}$ and $S_n^{(1)}$ are the corresponding first order 
excitation operators due to the interaction Hamiltonian. To calculate $\alpha_i(J_nM_n)$, only linear terms involving $T^{(1)}$ or $S_n^{(1)}$ operators are kept from the exponential function of Eqn. (\ref{eqn9}) 
\begin{eqnarray}
|\Psi_n \rangle &=& e^{T^{(0)}} \{ 1+ (1 + S_n^{(0)}) T^{(1)}+ S_n^{(1)} \} |\Phi_n\rangle .
\end{eqnarray}

Hence, the unperturbed and perturbed wave functions can be separated as
\begin{eqnarray}
|\Psi_n^{(0)} \rangle &=& e^{T^{(0)}} \{ 1 + S_n^{(0)} \} |\Phi_n\rangle , \\
|\Psi_n^{(1)} \rangle &=& e^{T^{(0)}} \{ (1 + S_n^{(0)}) T^{(1)}+ S_n^{(1)} \} |\Phi_n\rangle .
\label{eqn12}
\end{eqnarray}

 We consider only the single and double excitations from the RCC method (CCSD 
method) in our calculations, which is proved successful to cope with the 
electron correlation effects in most one valence systems; i.e.
\begin{eqnarray}
T &=& T_1 + T_2 \\
S_n &=& S_{1n} + S_{2n} .
\end{eqnarray}

 First we solve the unperturbed $T^{(0)}$ and $S_v^{(0)}$ amplitudes by
solving usual CC equations, then these amplitudes are used
to determine the $T^{(1)}$ and $S_v^{(1)}$ amplitudes in the following
equations
\begin{eqnarray}
\langle \Phi_0^{*} | \overline{\text{H}_N^{(\text{DC})}} T{(1)} | \Phi_0 \rangle =  -\langle \Phi_0^{*} | \overline{\text{H}_{\text{int}}} | \Phi_0 \rangle \\
\langle \Phi_n^{*} | (\overline{\text{H}_N^{(\text{DC})}} - \text{IP}_n) S_n^{(1)} | \Phi_n \rangle = \nonumber \\
- \langle \Phi_n^{*} | \left [ \overline{\text{H}_{\text{int}}} \{ 1 + S_n^{(0)}\} + \overline{\text{H}_N^{(\text{DC})}} T^{(1)} \right ] | \Phi_n \rangle ,
\end{eqnarray}
where the subscript $N$ represents normal order form of the operators, 
$\text{IP}_n$ is the ionization potential of the $n$th state and the symbol * represents
excited states with respect to the corresponding reference states.

After getting both the unperturbed and perturbed amplitudes, we evaluate 
polarizabilities using the following expression
\begin{widetext}
\begin{eqnarray}
\alpha_i(J_nM_n) &=& \frac {\langle \Phi_n | \{ 1 + S_n^{\dagger} \} e^{T^{\dagger}} O e^T \{ 1 + S_n \} | \Phi_n \rangle } { \langle \Phi_n | \{ 1 + S_n^{\dagger} \} e^{T^{\dagger}}  e^T \{ 1 + S_n \} | \Phi_n \rangle } \nonumber \\
&=& \frac {<\Phi_n|S_n^{(1)^{\dagger}}\overline{\text{O}^{(0)}} \{ 1+S_n^{(0)} \} + \{ 1+S_n^{(0)^{\dagger}} \} \overline{\text{O}^{(0)}}S_n^{(1)}+ \{ 1+ S_n^{(0)^{\dagger}} \} (T^{(1)^{\dagger}}\overline{\text{O}^{(0)}}+\overline{\text{O}^{(0)}}T^{(1)}) \{ 1 + S_n^{(0)} \} |\Phi_n>}{1+N_n^{(0)}} . \ \ \ \
\label{eqn19}
\end{eqnarray}
\end{widetext}
where for computational simplicity we define $\overline{\text{O}^{(0)}}=e^{T^{(0)^{\dagger}}} \text{O} e^{T^{(0)}}$  and $N_n^{(0)} = S_n^{(0)^{\dagger}} e^{T^{(0)^{\dagger}}} e^{T^{(0)}} S_n^{(0)}$. We compute $\overline{\text{O}^{(0)}}$ in two steps as effective one-body and two-body terms and substitute
in the above equation. We account for contributions from the normalization factor 
expressed as
\begin{eqnarray}
\text{Norm.} =  \langle \Psi_n | \text{O} | \Psi_n \rangle \{ \frac {1}{1+N_n} - 1 \}.
\label{eqn20}
\end{eqnarray}

 The above approach can easily be extended for a very precise calculations of  
PNC amplitudes 
\cite{wood} and CP violating electric dipole moments (EDMs) in atoms 
\cite{shukla} by obtaining the first order wavefunctions due
to nuclear spin independent and dependent weak interaction Hamiltonians in the 
place of dipole and quadrupole operators.
\begin{table}
\caption{Static dipole and quadrupole polarizabilities in alkali atoms: Li, Na and K.}
\begin{ruledtabular}
\begin{center}
\begin{tabular}{llll}
Atoms &  Expts & Others & This work \\
\hline\\
& \multicolumn{2}{c}{\underline{Dipole polarizabilities}} &  \\
Li & 164(3.4) \cite{molof} & 163.73 \cite{lim1}, 164.6 \cite{patil}   & 162.29 \\
 & 148(13) \cite{zorn} & 165.01 \cite{kobayashi}, 164.111 \cite{yan} &  \\
&  & &  \\
Na &  159.2(3.4) \cite{molof} & 163.07 \cite{safronova}, 164.89 \cite{lim1} & 162.89 \\
 & 164.6(11.5) \cite{zorn} & 160.7 \cite{patil}, 165.88 \cite{kobayashi} &  \\
 &  & &  \\
K & 292.8(6.1) \cite{molof} & 290.1 \cite{safronova}, 289.5 \cite{patil} & 286.01 \\
& 305(21.6) \cite{zorn} & 301.28 \cite{lim1},  285.23 \cite{kobayashi}  & \\
 &  & &  \\
& \multicolumn{2}{c}{\underline{Quadrupole polarizabilities}} &  \\
Li & & 1424(4) \cite{porsev}, 1393 \cite{patil} & 1421.28 \\
&  & 1423.266(5) \cite{yan}, 1424 \cite{marinescu} &  \\
&  & 1423 \cite{spelsberg} &  \\
&  & &  \\
Na &  & 1885(26) \cite{porsev}, 1796 \cite{patil} & 1899.67 \\
 &  & 1878 \cite{marinescu}, 1879 \cite{spelsberg} &  \\
 &  & &  \\
K &  & 5000(45) \cite{porsev}, 4703 \cite{patil} & 4919.71 \\
 &  & 5000 \cite{marinescu}, 5001 \cite{spelsberg} &  \\
\end{tabular}
\end{center}
\end{ruledtabular}
\label{tab1}
\end{table}

\begin{table}
\caption{Static dipole and quadrupole polarizabilities in alkaline earth-metal ions: Be$^+$, Mg$^+$ and Ca$^+$.}
\begin{ruledtabular}
\begin{center}
\begin{tabular}{llll}
Atoms &  Expts & Others & This work \\
\hline\\
& \multicolumn{2}{c}{\underline{Dipole polarizabilities}} &  \\
Be$^+$ &  & 24.93 \cite{patil}, 24.63 \cite{flannery} & 24.11 \\
 &  & 25.04 \cite{mukherjee}, 16.74 \cite{langhoff} &  \\
&  & &  \\
Mg$^+$ & 34.62(26) \cite{theodosiou} & 34.144 \cite{theodosiou}, 38.7 \cite{easa} & 34.59 \\
 & 33.0(5) \cite{chang} & 33.68 \cite{patil}, 37.2 \cite{langhoff} & \\
 & 33.8(8) \cite{lyons} & 34.0 \cite{adelman}, 38.9 \cite{kundu} &  \\
 &  & &  \\
Ca$^+$ & 70.89(15) \cite{theodosiou} & 70.872 \cite{theodosiou}, 112.4 \cite{easa} & 73.86 \\
 & 75.3(4) \cite{chang} & 71.01 \cite{patil}, 87(2) \cite{vaidyanathan} &  \\
 & 72.5(19) \cite{chang} & 76.9 \cite{adelman}, 96.2 \cite{langhoff} & \\
 &  & &  \\
& \multicolumn{2}{c}{\underline{Quadrupole polarizabilities}} &  \\
Be$^+$ &  & 52.93 \cite{patil}, 55.42 \cite{mukherjee} & 53.80 \\
&  & 55.71 \cite{langhoff}, 52.4 \cite{curtis} &  \\
&  & &  \\
Mg$^+$ &  & 150.2 \cite{patil}, 187.66 \cite{langhoff} & 156.17 \\
&  & 150.15 \cite{adelman} &  \\
 &  & &  \\
Ca$^+$ &  & 1171 \cite{patil}, 727.55 \cite{langhoff} & 706.59 \\
&  & 1303.51 \cite{adelman} &  \\
\end{tabular}
\end{center}
\end{ruledtabular}
\label{tab2}
\end{table}
\begin{table}
\caption{Contributions from DF and important perturbed CC terms for the dipole and quadrupole polarizabilities.}
\begin{ruledtabular}
\begin{center}
\begin{tabular}{lccccc}
Atoms &  DF & $\overline{O}S_{1n}^{(1)} + cc$ & $\overline{O}S_{2n}^{(1)} + cc$ & Norm. & Others \\
\hline\\
& \multicolumn{3}{c}{\underline{Dipole polarizabilities}} &  \\
Li & 168.95 & 164.21 & $-0.40$ & $-0.08$ & $-1.44$ \\
Na & 188.17 & 169.09 & $-1.26$ & $-0.29$ & $-4.65$ \\
K & 398.15 & 313.32 & $-6.93$ & $-2.11$ & $-18.28$ \\
Be$^+$ & 24.81 & 24.33 & $-0.08$ & $-0.01$ & $-0.13$ \\
Mg$^+$ & 38.39 & 35.57 & $-0.43$ & $-0.04$ & $-0.51$ \\
Ca$^+$ & 94.62 & 79.84 & $-1.48$ & $-0.32$ & $-4.18$ \\
&  & &  & \\
& \multicolumn{3}{c}{\underline{Quadrupole polarizabilities}} &  \\
Li & 1484.98 & 1444.87 & 0.00 & $-0.72$ & $-22.87$ \\
Na & 2230.62 & 2012.95 & 0.00 & $-3.34$ & $-109.94$ \\
K & 7099.70 & 5562.48  & 0.00 & $-36.38$ & $-606.39$  \\
Be$^+$ & 54.97 & 54.23 & 0.00 & $-0.02$ & $-0.41$ \\
Mg$^+$ & 171.34 & 161.09 & 0.00 & $-0.18$ & $-4.74$ \\
Ca$^+$ & 952.65 & 750.39 & 0.00 & $-3.07$ & $-40.73$ \\
\end{tabular}
\end{center}
\end{ruledtabular}
\label{tab3}
\end{table}
We construct the relativistic single particle orbitals using Gaussian type 
orbitals (GTOs) and we consider the finite size of the nucleus assuming a Fermi 
charge distribution as discussed in \cite{rajat}. One can use length and velocity gauge expressions for E1 and E2 operators to verify accuracy of the results.
However, we have used only the length gauge expressions which are more stable than others in our calculations. We present our results
in Tables \ref{tab1} and \ref{tab2}, and compare them with the experimental
results and other calculated values. Numerous calculations are available for the dipole polarizabilities in the neutral systems. However, we 
present only a few recent calculations in these tables. To our knowledge, 
there are only a few experimental results for the dipole polarizabilities 
 whereas no results are found for the quadrupole 
polarizabilities, though there exist a number of calculations.

\begin{figure}[h]
\includegraphics[width=8.5cm]{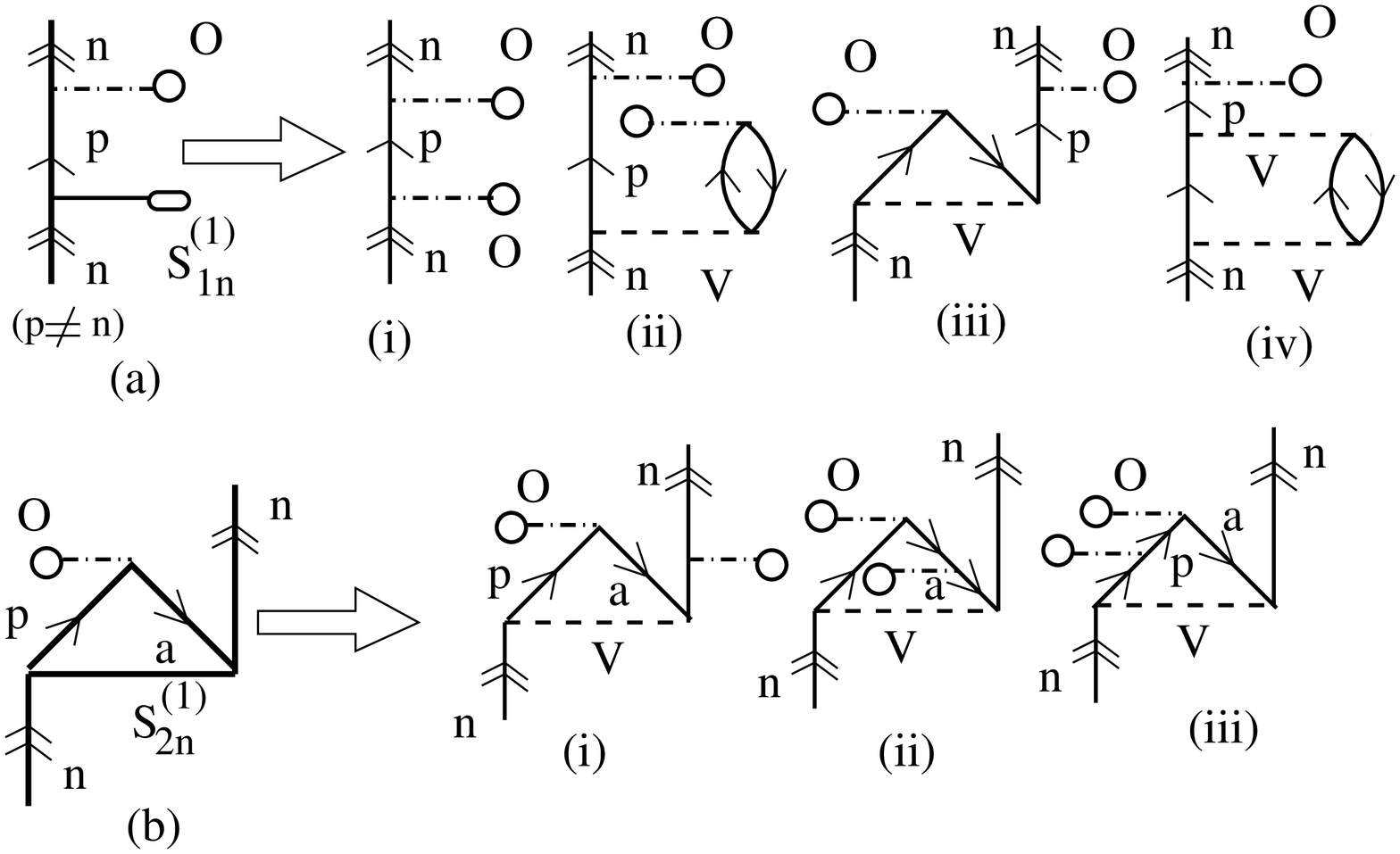}
\caption{Break down of important perturbed CC diagrams into some of the important lowest order MBPT diagrams. The $V$ represents Coulomb operator and a and p represent, occupied and unoccupied orbitals, respectively.}
\label{fig1}
\end{figure}
We present the DF and different RCC results in Table \ref{tab3} to understand 
the role of electron correlation effects.
As seen from this table, $OS_{1n}^{(1)}$ and its complex conjugate ($cc$) terms contribute predominantly. To interpret these contributions,
we break down this RCC term into some of the leading order MBPT diagrams
 as shown in the Fig. (\ref{fig1}(a)). 
It is obvious from this relationship
that the above term consists of dominant DF (Fig.\ref{fig1}(ai)), 
core-polarization (Fig.\ref{fig1}(aii) and \ref{fig1}(aiii)) and pair-correlation  (Fig.\ref{fig1}(aiv)) effects, therefore this results in large contributions  for
all the systems. However, we also show in the Fig. (\ref{fig1}(b)) another important RCC term  ($OS_{2n}^{(1)}$) as the sum of different types of core-polarization 
effects. From Table \ref{tab3}, we find that contributions from other higher order
terms are non-negligible while going towards the large systems. We also notice from this table that 
the trend of the correlation effects in neutral and ionic systems with the same electronic configurations are different.

 In conclusion, we have developed for the first time a general method in the relativistic coupled-cluster
theory to calculate the first order wave functions due to any rank operators
for both the parities. This technique can be applied to diverse areas of
physics ranging from polarizabilities to probe physics beyond the Standard Model of particle physics.
We also investigated electron correlation effects
from the dipole and quadrupole polarizabilities calculations in six different
systems and highlight their behavior. We observed that higher order 
correlation effects become important in large systems; suggesting that the
method can be employed rigorously in heavy systems to obtain accurate results,
and can be studied as {\it ab initio} tests. Indeed, this approach can also be
extended to determine frequency dependent (dynamic) polarizabilities 
which we defer to our next studies. Conclusively, the many-body aspects 
presented here would certainly be of interesting for both physicists and quantum chemists.

 We are grateful to Professor B. P. Das for many useful 
discussions and suggesting us to carry out this work. We are also delighted
to thank Professor P. Fulde for very enlightening discussions and 
hospitalities at MPI-PKS.

\end{document}